\documentclass{aastex61}

\usepackage{longtable}
\usepackage{multirow}
\received{\today}
\submitjournal{ApJ}

\shorttitle{An ALMA view of the N159W-South clump}
\shortauthors{Tokuda et al.}

\begin{document}

\title{An ALMA view of molecular filaments in the Large Magellanic Cloud. II. An early stage of high-mass star formation embedded at colliding clouds in N159W-South}

\correspondingauthor{Kazuki Tokuda}
\email{tokuda@p.s.osakafu-u.ac.jp}

\author[0000-0002-2062-1600]{Kazuki Tokuda}
\affiliation{Department of Physical Science, Graduate School of Science, Osaka Prefecture University, 1-1 Gakuen-cho, Naka-ku, Sakai, Osaka 599-8531, Japan}
\affiliation{National Astronomical Observatory of Japan, National Institutes of Natural Science, 2-21-1 Osawa, Mitaka, Tokyo 181-8588, Japan}

\author{Yasuo Fukui}
\affiliation{Department of Physics, Nagoya University, Chikusa-ku, Nagoya 464-8602, Japan}
\affiliation{Institute for Advanced Research, Nagoya University, Furo-cho, Chikusa-ku, Nagoya 464-8601, Japan}

\author{Ryohei Harada}
\affiliation{Department of Physical Science, Graduate School of Science, Osaka Prefecture University, 1-1 Gakuen-cho, Naka-ku, Sakai, Osaka 599-8531, Japan}

\author{Kazuya Saigo}
\affiliation{National Astronomical Observatory of Japan, National Institutes of Natural Science, 2-21-1 Osawa, Mitaka, Tokyo 181-8588, Japan}

\author{Kengo Tachihara}
\affiliation{Department of Physics, Nagoya University, Chikusa-ku, Nagoya 464-8602, Japan}

\author{Kisetsu Tsuge}
\affiliation{Department of Physics, Nagoya University, Chikusa-ku, Nagoya 464-8602, Japan}

\author{Tsuyoshi Inoue}
\affiliation{Department of Physics, Nagoya University, Chikusa-ku, Nagoya 464-8602, Japan}

\author{Kazufumi Torii}
\affiliation{Nobeyama Radio Observatory, 462-2 Nobeyama Minamimaki-mura, Minamisaku-gun, Nagano 384-1305, Japan}

\author{Atsushi Nishimura}
\affiliation{Department of Physical Science, Graduate School of Science, Osaka Prefecture University, 1-1 Gakuen-cho, Naka-ku, Sakai, Osaka 599-8531, Japan}

\author{Sarolta Zahorecz}
\affiliation{Department of Physical Science, Graduate School of Science, Osaka Prefecture University, 1-1 Gakuen-cho, Naka-ku, Sakai, Osaka 599-8531, Japan}
\affiliation{National Astronomical Observatory of Japan, National Institutes of Natural Science, 2-21-1 Osawa, Mitaka, Tokyo 181-8588, Japan}

\author{Omnarayani Nayak}
\affiliation{Space Telescope Science Institute, 3700 San Martin Drive, Baltimore, MD 21218, USA}

\author{Margaret Meixner}
\affiliation{Department of Physics \& Astronomy, Johns Hopkins University, 3400 N. Charles Street, Baltimore, MD 21218, USA}
\affiliation{Space Telescope Science Institute, 3700 San Martin Drive, Baltimore, MD 21218, USA}

\author{Tetsuhiro Minamidani}
\affiliation{Nobeyama Radio Observatory, 462-2 Nobeyama Minamimaki-mura, Minamisaku-gun, Nagano 384-1305, Japan}

\author{Akiko Kawamura}
\affiliation{National Astronomical Observatory of Japan, National Institutes of Natural Science, 2-21-1 Osawa, Mitaka, Tokyo 181-8588, Japan}

\author{Norikazu Mizuno}
\affiliation{National Astronomical Observatory of Japan, National Institutes of Natural Science, 2-21-1 Osawa, Mitaka, Tokyo 181-8588, Japan}
\affiliation{Department of Astronomy, School of Science, The University of Tokyo, 7-3-1 Hongo, Bunkyo-ku, Tokyo 133-0033, Japan}

\author{Remy Indebetouw}
\affiliation{Department of Astronomy, University of Virginia, P.O. Box 400325, Charlottesville, VA 22904, USA}
\affiliation{National Radio Astronomy Observatory, 520 Edgemont Road, Charlottesville, VA 22903, USA}

\author{Marta Sewi{\l}o}
\affiliation{CRESST II and Exoplanets and Stellar Astrophysics Laboratory, NASA, Goddard Space Flight Center, Greenbelt, MD 20771, USA}
\affiliation{Department of Astronomy, University of Maryland, College Park, MD 20742, USA}

\author{Suzanne Madden}
\affiliation{AIM, CEA, CNRS, Universit\'e Paris-Saclay, Universit\'e Paris Diderot, Sorbonne Paris Cit\'e, F-91191 Gif-sur-Yvette, France}

\author{Maud Galametz}
\affiliation{AIM, CEA, CNRS, Universit\'e Paris-Saclay, Universit\'e Paris Diderot, Sorbonne Paris Cit\'e, F-91191 Gif-sur-Yvette, France}

\author{Vianney Lebouteiller}
\affiliation{AIM, CEA, CNRS, Universit\'e Paris-Saclay, Universit\'e Paris Diderot, Sorbonne Paris Cit\'e, F-91191 Gif-sur-Yvette, France}

\author{C.-H. Rosie Chen}
\affiliation{Max Planck Institute for Radio Astronomy, Auf dem Huegel 69, D-53121 Bonn, Germany}

\author{Toshikazu Onishi}
\affiliation{Department of Physical Science, Graduate School of Science, Osaka Prefecture University, 1-1 Gakuen-cho, Naka-ku, Sakai, Osaka 599-8531, Japan}



\begin{abstract}
We have conducted ALMA CO isotopes and 1.3\,mm continuum observations toward filamentary molecular clouds of the N159W-South region in the Large Magellanic Cloud with an angular resolution of $\sim$0\farcs25 ($\sim$0.07\,pc). Although the previous lower-resolution ($\sim$1\arcsec) ALMA observations revealed that there is a high-mass protostellar object at an intersection of two line-shaped filaments in $^{13}$CO with the length scale of $\sim$10\,pc, the spatially resolved observations, in particular, toward the highest column density part traced by the 1.3\,mm continuum emission, the N159W-South clump, show complicated hub-filamentary structures. We also discovered that there are multiple protostellar sources with bipolar outflows along the massive filament. The redshifted/blueshifted components of the $^{13}$CO emission around the massive filaments/protostars have complementary distributions, which is considered to be a possible piece of evidence for a cloud--cloud collision. We propose a new scenario in which the supersonically colliding gas flow triggers the formation of both the massive filament and protostars. This is a modification of the earlier scenario of cloud--cloud collision, by Fukui et al., that postulated the two filamentary clouds occur prior to the high-mass star formation. A recent theoretical study of the shock compression in colliding molecular flows by Inoue et al. demonstrates that the formation of filaments with hub structure is a usual outcome of the collision, lending support for the present scenario. The theory argues that the filaments are formed as dense parts in a shock compressed sheet-like layer, which resembles $``$an umbrella with pokes.$"$
\end{abstract}

\keywords{stars: formation  --- stars: protostars --- ISM: clouds--- ISM:  kinematics and dynamics --- ISM: individual (N159W)}

\section{Introduction} \label{sec:intro}
Though high-mass stars are considered to have great impacts on the galaxy evolution, their formation process is not fully understood. There are a number of works that investigated the high-mass star formation mechanism (for reviews, e.g., \citealt{Zinnecker07,Tan14}). Although the precursors of massive stars are supposed to be very dense and massive ($\sim$100\,$M_{\odot}$) cores \citep[e.g.,][]{Krumholz09,Peretto13}, it may be a problem how such peculiar cores are formed as an initial condition of massive star formation. \cite{Habe92} suggested that molecular cloud collisions form a dense shock compressed layer, which is massive enough to form a massive star based on their numerical simulations in a short timescale less than 1\,Myr (see also, \citealt{Takahira14,Matsumoto15,Shima18}). \cite{Inoue18} demonstrated that shock compressions induced by cloud--cloud collision promote massive filament formation, which is perpendicular to the background magnetic field (see also \citealt{Inoue13}). O-type stars are formed by the global collapse of the massive filament with a high mass-accretion rate, $>$10$^{-4}$\,$M_{\odot}$\,yr$^{-1}$. 

Because the high-mass protostars are supposed to be formed in giant molecular clouds (GMCs), detailed observational studies toward the GMCs are thus needed to examine the initial conditions of high-mass star formation. Large-scale surveys with high-angular resolution ($<$0.1--1\,pc) in the Galaxy have been providing us with fruitful knowledge on the physical properties of molecular clouds and (high-mass) star formation. Recent high-angular resolution observations with ground-based single-dish telescopes and the {\it Herschel} satellite revealed that filamentary structures are ubiquitous in both dark clouds and GMCs \citep[e.g.,][]{Mizuno95,Nagahama98,Onishi96,Onishi99,Doris11,Doris19,Andre14,Andre16}. High-mass young stellar objects (YSOs) tend to be located at the intersection of multiple filamentary clouds, called $``$hub filament$"$ \citep[e.g.,][]{Myers09,Peretto13}. The central part of hub filaments is as massive as $\gtrsim$1000\,$M_{\odot}$\,pc$^{-1}$, which is sometimes referred to as a $``$ridge$"$ \citep[e.g.,][]{Motte07,Hill12,Nguyen13}. \cite{Motte18} and references therein suggested that such massive filaments are supposed to be products of the global hierarchical collapse of molecular clouds. Because the line mass of the ridges is significantly higher than the critical line mass of an isothermal filament \citep[see][]{Inutsuka97}, by up to two orders of magnitude, they are supposed to be unstable against the global collapse and fragmentation. Although some formation/stabilization mechanism of the hub filaments and the ridges, such as large-scale compression and internal MHD (magneto-hydrodynamic) waves, have been proposed \citep{Andre16}, the true nature is not fully understood, possibly due to the lack of the suitable targets in the solar neighborhood.

ALMA is capable of resolving internal structures of molecular clouds even in external galaxies. In particular, the Large Magellanic Cloud (LMC) is an ideal laboratory to investigate high-mass star formation thanks to its nearly face-on view \citep{Balbinot15} and the close distance, $\sim$50\,kpc \citep{Schaefer08,de14}. It is also a great advantage to directly compare the distributions of molecular gas observed by ALMA and positions of massive YSOs identified by {\it Spitzer} and {\it Herschel} \citep[e.g.,][]{Gruendl09,Chen10,Seale14} without any serious contamination in the line of sight. Earlier studies using the H$\;${\sc i} gas observations by \cite{Fukui17} found that there are supergiant shells \citep{Kim99,Kim03} and kiloparsec-scale gas flows caused by the last tidal interaction between the LMC and the Small Magellanic Clouds (SMC). Therefore, we may be able to examine the relation between such large-scale gas kinematics and the local star formation activities. 
 Our present target in this paper is the N159W-South clump, which was discovered by our previous ALMA Cycle 1 observations (\citealt{Fukui15}, hereafter Paper I) with an angular resolution of $\sim$1\arcsec ($\sim$0.24\,pc) toward a GMC in the N159W region \citep[e.g.,][]{Johansson98,Minamidani08,Minamidani11}. Paper I revealed that the GMC is composed of many filamentary molecular clouds and discovered the first example of protostellar outflows in the external galaxies. Paper I also found that the protostellar source with a stellar mass of $\sim$37\,$M_{\odot}$ in the N159W-South clump is located toward an intersection of two filaments and suggested that the filament--filament collision triggered the protostar formation. Although the ALMA observations significantly improved our understanding of molecular cloud structures and star formation in this object, much higher-angular-resolution studies are needed to further resolve the filamentary structures down to a width of $\lesssim$0.1\,pc \citep[see,][]{Doris11,Doris19} and investigate the star formation activities therein. 
 
In this paper, we present high-angular resolution observations obtained in ALMA Cycle 4 (P.I.: Y. Fukui \#2016.1.01173.S) with $\sim$0\farcs25 ($\sim$0.07\,pc) resolution toward the N159W-South clump. We also observed the N159W-North and the N159E-Papillon region in the same project. The observational results of the N159E-Papillon region, which is considered to be in a later evolutionary stage than the N159W-South clump \citep{Saigo17}, are presented in a separate paper (\citealt{Fukui19}, hereafter FTS19). 

\section{Observations} \label{sec:Obs}
We carried out ALMA Cycle 4 Band 6 (211--275 GHz) observations toward the N159W with the main array 12\,m antennas. The observations of N159W-South centered at ($\alpha_{\rm J2000.0}$, $\delta_{\rm J2000.0}$) = (5$^{\rm h}$39$^{\rm m}$41\fs0, $-$69\arcdeg46\arcmin06\farcs0) were carried out between 2016 November and 2017 July. There were three spectral windows targeting $^{12}$CO\,($J$ = 2--1), $^{13}$CO\,($J$ = 2--1) and C$^{18}$O\,($J$ = 2--1) with a bandwidth of 58.6\,MHz. The frequency resolutions were 30.6\,kHz for $^{12}$CO\,($J$ = 2--1) and 61.0\,kHz for the others. We used two spectral windows for the continuum observations with the aggregated bandwidth of 3.75\,GHz. The observed frequencies include some line emission, such as the radio recombination line of H30$\alpha$ and SiO ($J$ = 5--4). The projected baseline length ranges from 14 to 1940\,m. 

The data were processed with the CASA (Common Astronomy Software Application) package \citep{McMullin07} version 5.0.0. We used the \texttt{tclean} task in the imaging process with the \texttt{multi-scale} deconvolver to recover the extended emission. The imaging grid and velocity channels were 0\farcs07 and 0.2\,km\,s$^{-1}$, respectively, and we applied the Briggs weighting with the robust parameter of 0.5. We used the \texttt{auto-multithresh} procedure in \texttt{tclean} to select the emission mask in the dirty and residual images. We continued the deconvolution process until the intensity of the residual image reached $\sim$1$\sigma$ noise level. The synthesized beams of the continuum and the $^{13}$CO\,($J$ = 2--1) observations are 0\farcs26 $\times$ 0\farcs23 and 0\farcs29 $\times$ 0\farcs25, respectively. The (1$\sigma$) RMS noise levels of the line and the 1.3\,mm continuum are $\sim$4.5\,mJy\,beam$^{-1}$ ($\sim$1.5 K) at a velocity resolution of 0.2\,km\,s$^{-1}$ and $\sim$0.027\,mJy\,beam$^{-1}$. We concluded that the previous ALMA Cycle 1 observations (Paper I, see also \citealt{Nayak18}) did not show significant missing flux of this source based on a comparison between the ALMA and the single-dish observations. The total fluxes of the $^{12}$CO, $^{13}$CO, and the 1.3\,mm continuum data of the Cycle 4 were $\sim$20\% lower than those of the Cycle 1. Because the current observations fully cover the baseline range of the previous one, 16--395\,m, the discrepancies are supposed to be mainly caused by the calibration error instead of the missing flux. We thus use the Cycle 4 data alone in the following analysis.

\section{Results \label{Results}} 
\subsection{A dense filamentary cloud traced by the 1.3\,mm dust-continuum emission associated with molecular outflows \label{R:outflow}}
\begin{figure}[tbp]
\includegraphics[width=180mm]{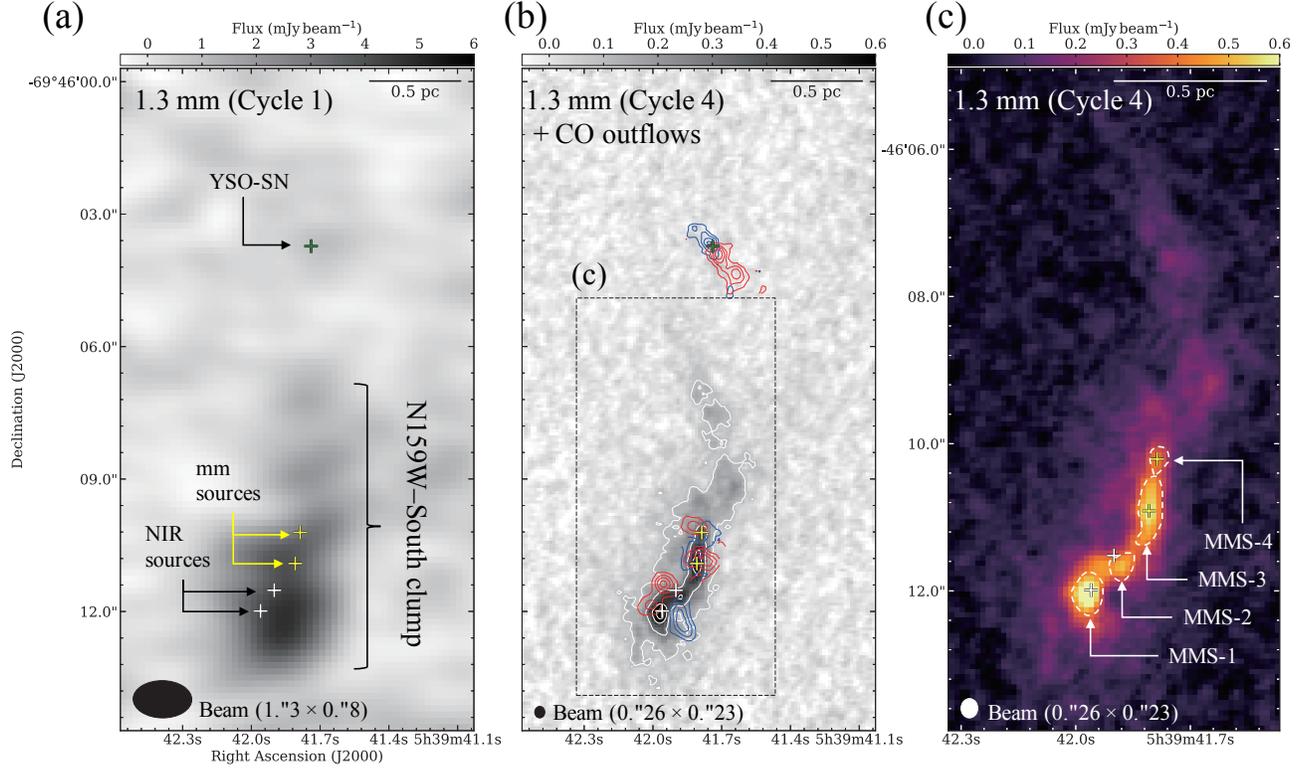}
\caption{1.3\,mm continuum distributions and molecular outflows toward the N159W-South clump. (a) Gray-scale image of the 1.3\,mm continuum emission obtained in the previous Cycle 1 study (Paper I). The white crosses represent the positions of the infrared sources identified with the VLT observations \citep{Testor06}. The yellow and green crosses denote the dust-continuum peaks of MMS-3 and 4, and the median position between the redshifted and blueshifted outflow lobes shown in panel (b), respectively. (b) The gray-scale image and white contours are the same as (a) but for the high-resolution Cycle 4 data. The minimum contour level and the subsequent steps are 0.1 and 0.2\,mJy\,beam$^{-1}$, respectively.
Redshifted and blueshifted outflow lobes in $^{12}$CO\,($J$ = 2--1) are shown in blue and red contours, respectively. The contour levels are 10, 20, 30, and 40\,K\,km\,s$^{-1}$. The integrated-velocity ranges are shown in dashed lines in panels (a-d) of Figure \ref{fig:outspect}. (c) Zoomed-in view of the N159W-South clump in 1.3\,mm continuum showing the filamentary structure. The white dashed lines, corresponding to $\sim$0.4\,mJy\,beam$^{-1}$, represent the boundary of the identified millimeter sources.
\label{fig:outflow}}
\end{figure}

\ Figures \ref{fig:outflow} (a-c) show 1.3\,mm dust continuum distributions toward the N159W-South clump. Although the previous low-resolution ($\sim$1\arcsec) data shown in the panel (a) identified the clump as a single-source with a north--south elongation (see also Paper I), the present Cycle 4 high-resolution observations clearly resolved the filamentary structure with multiple local maxima. We named the four major continuum maxima as MMS-1, 2, 3, and 4 labeled in Figure \ref{fig:outflow} (c).
The positions of MMS-1 and MMS-2 correspond to near-infrared (NIR) sources, \#121 and \#123, respectively \citep{Testor06}.
These sources were originally discovered as a single source (P2) by \cite{Jones86}, and the subsequent high-resolution observations with the VLT (Very Large Telescope) resolved it into more than two sources (\citealt{Testor06}, see also the Gemini observations by \citealt{Bernard16}).
In the $K_{\rm s}$-band observations, the two NIR sources have almost the same magnitude (13.59\,mag for \#121, and 13.40\,mag for \#123), and are the brightest ones and highly reddened among the Testor's sample. \cite{Testor06} concluded that the two sources are considered to be YSOs. MMS-3 and 4 have no counterparts in the infrared observations. The sources detected by \cite{Testor06} were in the $K_{\rm s}$ magnitude range of 13.40--19.4\,mag. Although this means that the embedded sources in MMS-3 and 4 are much fainter than the known NIR sources, the presence of the outflow activities (see the later paragraph) strongly indicates that there are YSOs in the millimeter-sources. We derived the physical properties of the millimeter-sources as listed in Table \ref{table:1.3mm}. We assumed the absorption coefficient per unit dust mass at 1.3\,mm, and the dust-to-gas mass ratio and the dust temperature to be 1\,cm$^{2}$\,g$^{-1}$, 3.5 $\times$ 10$^{-3}$, and 20\,K, respectively, to derive the gas mass from the dust emission (see, \citealt{Oss94}; \citealt{Herrera13}; \citealt{Gordon14}; Paper I). The separations among the dust peaks are roughly $\sim$0.2\,pc.

We characterized the properties of the 1.3\,mm filamentary clouds.
The total mass of the filamentary structure traced by the continuum above the 3$\sigma$ detection is $\sim$4 $\times$ 10$^3$\,$M_{\odot}$, which is consistent with the previous estimate (Paper I).
We searched the peak positions (i.e., crests of the filament) in the R.A. direction along each decl. bin of Figure \ref{fig:outflow} (c). The length of the filament crests is $\sim$1.7\,pc. We performed the Gaussian fitting to the cross section taken along the filament crest. The averaged width (FWHM) of the filament is $\sim$0.14\,pc. The resultant line mass (=total mass/length) of the filament is $\sim$2 $\times$ 10$^3$\,$M_{\odot}$\,pc$^{-1}$. These properties are consistent with those of $``$ridges$"$ in Galactic high-mass star-forming clouds (e.g., NGC\,6334, \citealt{Andre16}). 

\begin{deluxetable}
{cccccccccc}  
\tabletypesize{\scriptsize}
\tablecaption{Properties of 1.3\,mm continuum sources in the N159W-South clump \label{table:1.3mm}}
\tablewidth{0pt}
\tablehead{
Name & $\alpha {\rm (J2000.0)}$ & $\delta {\rm (J2000.0)}$ & $F_{\nu}$ (mJy)$^{\rm a}$ & $F_{\rm max}$ (mJy\,beam$^{-1}$) & $N_{\rm peak}$ (cm$^{-2}$)$^{\rm b}$ & $M$ ($M_{\odot}$)$^{\rm c}$ & NIR source$^{\rm d}$}
\startdata
MMS-1 & 5$^{\rm h}$39$^{\rm m}$41\fs97 & -69\arcdeg46\arcmin12\farcs05 & 1.5 & 0.82 & 1.2\,$\times$\,10$^{24}$ & 2.1\,$\times$\,10$^{2}$ & 121\\
MMS-2 & 5$^{\rm h}$39$^{\rm m}$41\fs88 & -69\arcdeg46\arcmin11\farcs69 & 0.47 & 0.52 & 8.5\,$\times$\,10$^{23}$ & 6.2\,$\times$\,10$^{1}$ & 123\\
MMS-3 & 5$^{\rm h}$39$^{\rm m}$41\fs81 & -69\arcdeg46\arcmin10\farcs93 & 1.6 & 0.63 & 1.0\,$\times$\,10$^{24}$ & 2.1\,$\times$\,10$^{2}$ & $\cdots$ \\
MMS-4 & 5$^{\rm h}$39$^{\rm m}$41\fs79 & -69\arcdeg46\arcmin10\farcs25 & 0.43 & 0.41 & 6.5\,$\times$\,10$^{23}$ & 5.6\,$\times$\,10$^{1}$ & $\cdots$\\
\enddata
\ \ 
\tablenotetext{\rm a}{Flux of the millimeter emission integrated above the white dashed lines in each source.}
\tablenotetext{\rm b}{H$_2$ column density at the peak position.}
\tablenotetext{\rm c}{Total mass integrated above the white dashed lines in each source.}
\tablenotetext{\rm d}{Name of associated near infrared sources \citep{Testor06}.}
\end{deluxetable}

\begin{figure}[htbp]
\begin{center}
\includegraphics[width=140mm,angle=-90]{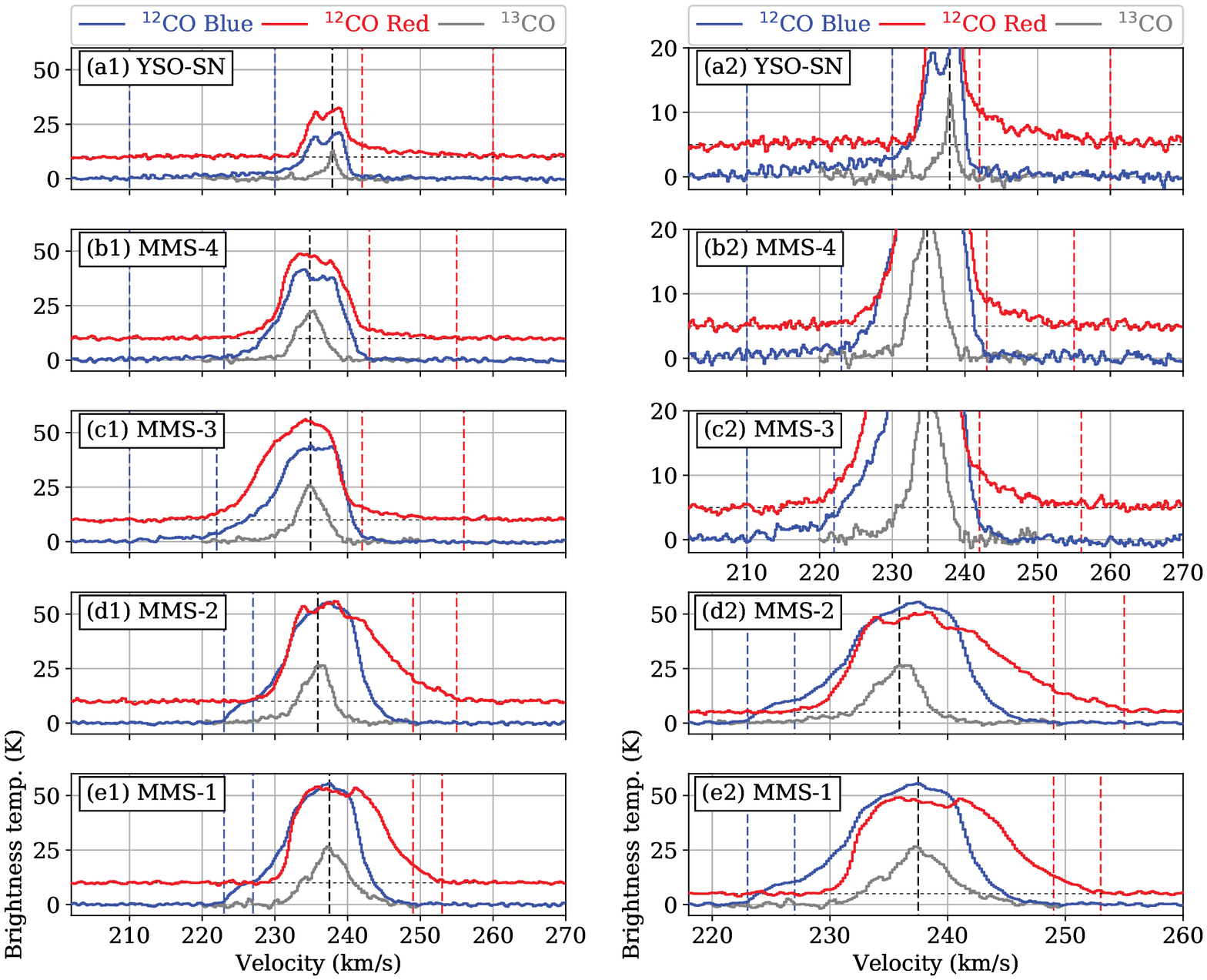}
\caption{Outflow spectra toward the N159W-South clump. (a1)--(d1) Red and blue lines of averaged spectra of the outflow wings for MMS-1,2,3,4 and YSO-SN over the regions inside the lowest red and blue contours, respectively, in Figure \ref{fig:outflow} (b). The blue and red dotted lines show the defined velocity ranges of the outflows (see also Table \ref{table:Outflow}). Note that the same spectra are plotted as the blue wings in panels (d1) and (e1) (see the text). The red lines are offset by $+$10\,K for the visualization. The gray lines show the $^{13}$CO\,($J$ = 2--1) spectra at the positions of MMS-1, 2, 3 and YSO-SN. Black lines indicate the central velocities obtained by fitting the $^{13}$CO spectra with a single Gaussian profile. (a2)--(e2) Same as (a1)--(e1) but for the enlarged views (in Y-axis for (a)--(c), and X-axis for (d) and (e)), to stress the wing features of the outflow spectra. 
\label{fig:outspect}}
\end{center}
\end{figure}

We have detected compact high-velocity wings in the $^{12}$CO\,($J$ = 2--1) observations tracing outflows from the protostellar sources. 
We identified the outflow components as follows. Based on the $^{12}$CO channel map, we searched high-velocity emission more than 10\,km\,s$^{-1}$ apart from the systemic velocity ($\sim$236\,km\,s$^{-1}$) by eye, and then we extracted the spectra of the high-velocity emission. We manually selected the lowest velocities of the redshifted/blueshifted outflows to avoid the strong emission of the parental filamentary clouds. Using the velocity-smoothed data, we defined the highest velocities where the intensities of the wing components are close to zero levels. The red and blue contours in Figure \ref{fig:outflow} (b) show the distributions of the outflow components, and the averaged spectra in each outflow are shown in Figure \ref{fig:outspect}. For the sources in Figure \ref{fig:outflow} (c), the previous $^{12}$CO observations identified these multiple flows as a single bipolar flow due to the lack of the angular resolution (Figure 2 in Paper I). The blue/red wings from MMS-1 and 2 are possibly merged, as shown in Figure \ref{fig:outflow} (b).  Especially for the blue component, it is hard to separate the individual components. We plotted the same spectra in the Figures \ref{fig:outspect} (d), (e) and listed as the blue wing of MMS-1. Although the redshifted components around the MMS-1 and 2 are also spatially merging, there are multiple local peaks in the contours. The presence of at least two individual YSOs (\#121, and \#123) in each millimeter source indicate that the spatially merging blue/redshifted lobes are originated from the protostars. Nevertheless, higher-angular-resolution observations in CO will be able to clearly resolve the individual outflow components. On the other hand, the outflows around MMS-2 and 3 are well spatially separated, indicating that they are individual outflows launched by the embedded YSOs. Note that the broadening of the $^{12}$CO linewidths toward the millimeter sources is considered to be originate from not only the outflowing gas but from cloud--cloud collision. We discuss the latter effect in Sect. \ref{R:filament}.

In addition to the above, we discovered a new candidate of bipolar outflow $\sim$2\,pc away from the N159W-South clump (Figures \ref{fig:outflow}(b) and \ref{fig:outspect}(a)). The green crosses indicate the position of the center of the high-velocity gas in Figures \ref{fig:outflow} (a,b). This strongly indicates that there is at least one embedded protostar; we hereafter call this source YSO-SN. Although this source is located in the $^{13}$CO filamentary cloud, as shown in Figure \ref{fig:13COimage}, we could not detect any 1.3\,mm continuum emission with the present sensitivity. This suggests that the parental core is less massive than the other millimeter sources in the N159W-South clump. We need high-sensitivity and high-angular resolution observations in millimeter/submillimeter continuum or high-density molecular gas tracers to characterize the nature of the parental core of YSO-SN.

We estimated the physical parameters of these outflows and have listed them in Table \ref{table:Outflow}. In order to derive the masses of the outflows, we adopted a conversion factor from $^{12}$CO\,($J$ = 1--0) intensity to the column density of $X_{\rm CO}$ = 7 $\times$ 10$^{20}$\,cm$^{-2}$ (K\,km\,s$^{-1}$) and the $^{12}$CO\,($J$ = 2--1)/$^{12}$CO\,($J$ = 1--0) ratio = 1.0, with the case of a gas kinematic temperature of $\sim$30\,K and a volume density of $\sim$10$^4$\,cm$^{-3}$ \citep{Beuther02b}. We measured the projected distances to the peak intensity of the outflow lobes from the continuum peaks MMS-1, 2, 3, and 4. For YSO-SN, we adopted the length of the blueshifted and redshifted lobes as the distance. Using the maximum velocity with respect to the systemic velocity measured from the $^{13}$CO spectra (Figure \ref{fig:outspect}), we estimated the dynamical time ($t_{d}$ = distance/velocity) of the flow by assuming an inclination angle of 30$\arcdeg$--70$\arcdeg$.
We calculated the outflow force ($F_{\rm CO}$) by using the procedure in \cite{Beuther02b}. The relations between the $F_{\rm CO}$ and the core mass derived from the dust emission for MMS-1, 2, 3 and 4 roughly follow those seen at massive protostellar sources in the Galaxy \citep{Beuther02b}. Although we could not detect dust emission from YSO-SN, this is also considered to be a massive source because the outflow parameters are very similar to those of the other massive sources. We thus conclude that at least five protostellar sources with separations of $\sim$0.2--2\,pc are forming along the filamentary cloud without significant time delay more than $\sim$10$^4$ yr based on the dynamical time of the outflows (Table \ref{table:Outflow}). We also found similar star formation activities in the N159E-Papillon region (FTS19). Note that the directions of the outflows in the N159W-South clump are roughly perpendicular to the orientation of the filament, indicating that the magnetic field directions are also perpendicular to that. We discuss the formation scenario of the YSOs and the filament in Sect. \ref{Dis}. 

\begin{deluxetable}
{ccccccc}  
\tabletypesize{\scriptsize}
\tablecaption{Outflow properties in the N159W-South clump\label{table:Outflow}}
\tablewidth{0pt}
\tablehead{
Source Name & Outflow Lobe  & Mass ($M_{\odot})$ & Distance (pc) & Velocity$^{\rm a}$ (km\,s$^{-1}$) & Velocity Range (km\,s$^{-1}$) & $t_{\rm b}$$^{\rm b}$ (yr) 
}
\startdata
\multirow{2}{*}{MMS-1} & Blue lobe$^{\rm c}$  & 7.2 & 0.13 & 14.5 & 223--227 & 5.1--25 $\times$ 10$^{3}$ \\
                                      & Red lobe  & 2.8 & 0.07 & 15.5 & 249--253 & 2.6--13 $\times$ 10$^{3}$ \\
\hline
MMS-2 & Red lobe        & 3.4 & 0.12 & 19.1 & 249--255 & 3.6--17 $\times$ 10$^{3}$ \\
\hline 
\multirow{2}{*}{MMS-3} & Blue lobe  & 4.4 & 0.06 & 24.9 & 210--222 & 1.4--6.5 $\times$ 10$^{3}$ \\
                                     & Red lobe  & 4.5 & 0.05 & 21.1 & 242--256 & 1.3--6.1 $\times$ 10$^{3}$ \\
\hline 
\multirow{2}{*}{MMS-4} & Blue lobe  & 1.2 & 0.05 & 24.8 & 210--223 & 1.1--5.4 $\times$ 10$^{3}$ \\
                                     & Red lobe  & 2.6 & 0.06 & 20.2 & 243--255 & 1.7--8.0 $\times$ 10$^{3}$ \\
\hline 
\multirow{2}{*}{YSO-SN} & Blue lobe  & 1.5 & 0.17 & 22.1 & 210--230 & 4.3--20 $\times$ 10$^{3}$ \\
                                        & Red lobe  & 3.2 & 0.17 & 27.9 & 242--260 & 3.4--16 $\times$ 10$^{3}$ \\
\enddata
\tablenotetext{\rm a}{Maximum radial velocity of the outflow lobe with respect to the systemic velocity.}
\tablenotetext{\rm b}{Dynamical time assuming the inclination angles of 30$\arcdeg$--70$\arcdeg$.}
\tablenotetext{\rm c}{This blue lobe is considered to be merged with that from MMS-2 (see the text and Figure \ref{fig:outflow} (c)).}
\end{deluxetable}

\subsection{Velocity and spatial structures of the $^{13}$CO filamentary clouds \label{R:filament}}
\begin{figure}[htbp]
\begin{center}
\includegraphics[width=200mm]{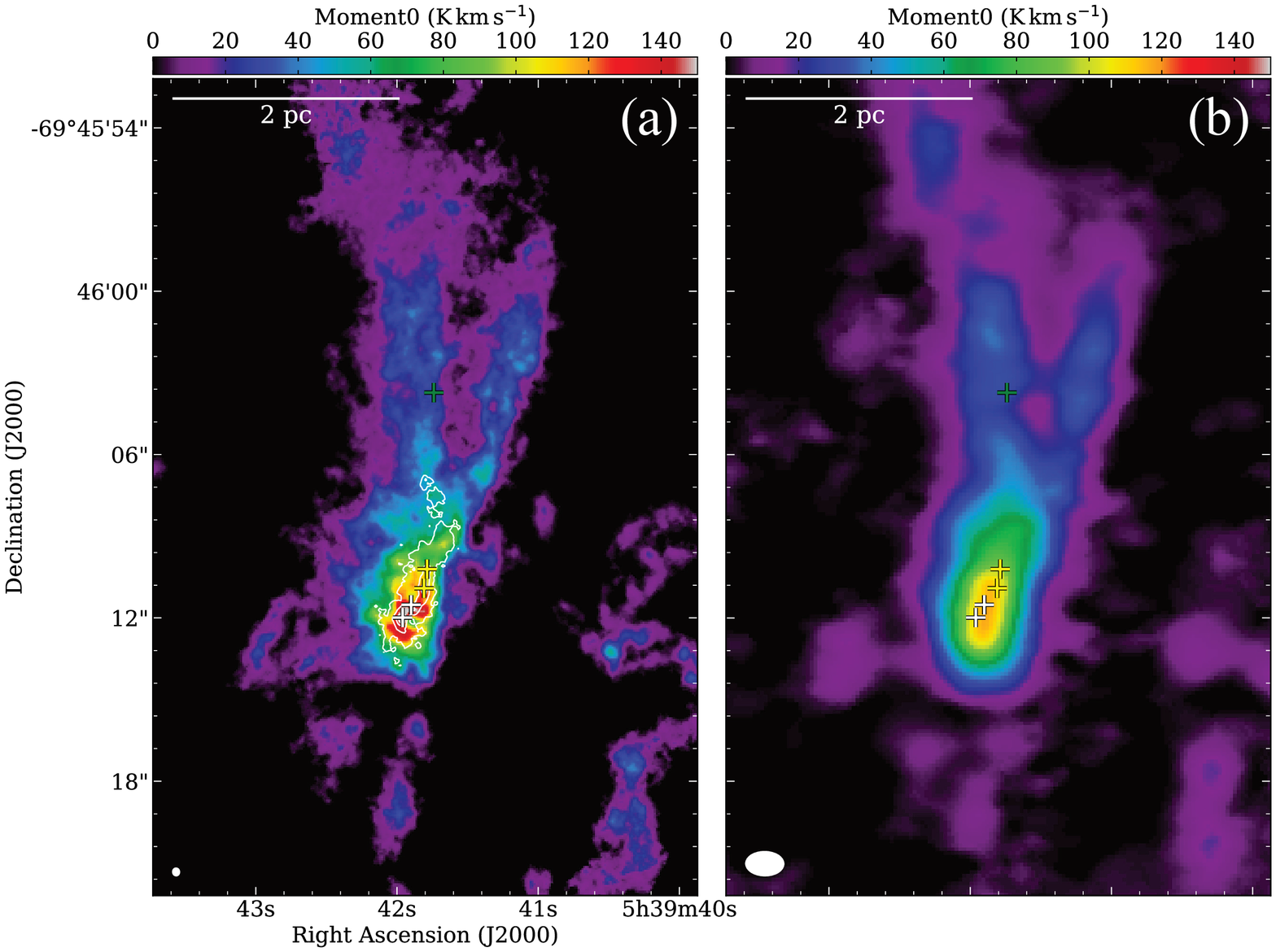}
\caption{$^{13}$CO\,($J$ = 2--1) distributions of the N159W-South region. (a) Color scale of the velocity-integrated intensity map of $^{13}$CO\,($J$ = 2--1) obtained by the Cycle 4 observations. The white, yellow, and green crosses and white  contours are the same as those in Figure \ref{fig:outflow}. The angular resolution, 0\farcs29 $\times$ 0\farcs25, is given by the white ellipse in the lower left corner. (b) Same as (a) but for the Cycle 1 data. The angular resolution, 1\farcs3 $\times$ 0\farcs8, is given by the white ellipse in the lower left corner.
\label{fig:13COimage}}
\end{center}
\end{figure}

\begin{figure}[htbp]
\begin{center}
\includegraphics[width=190mm]{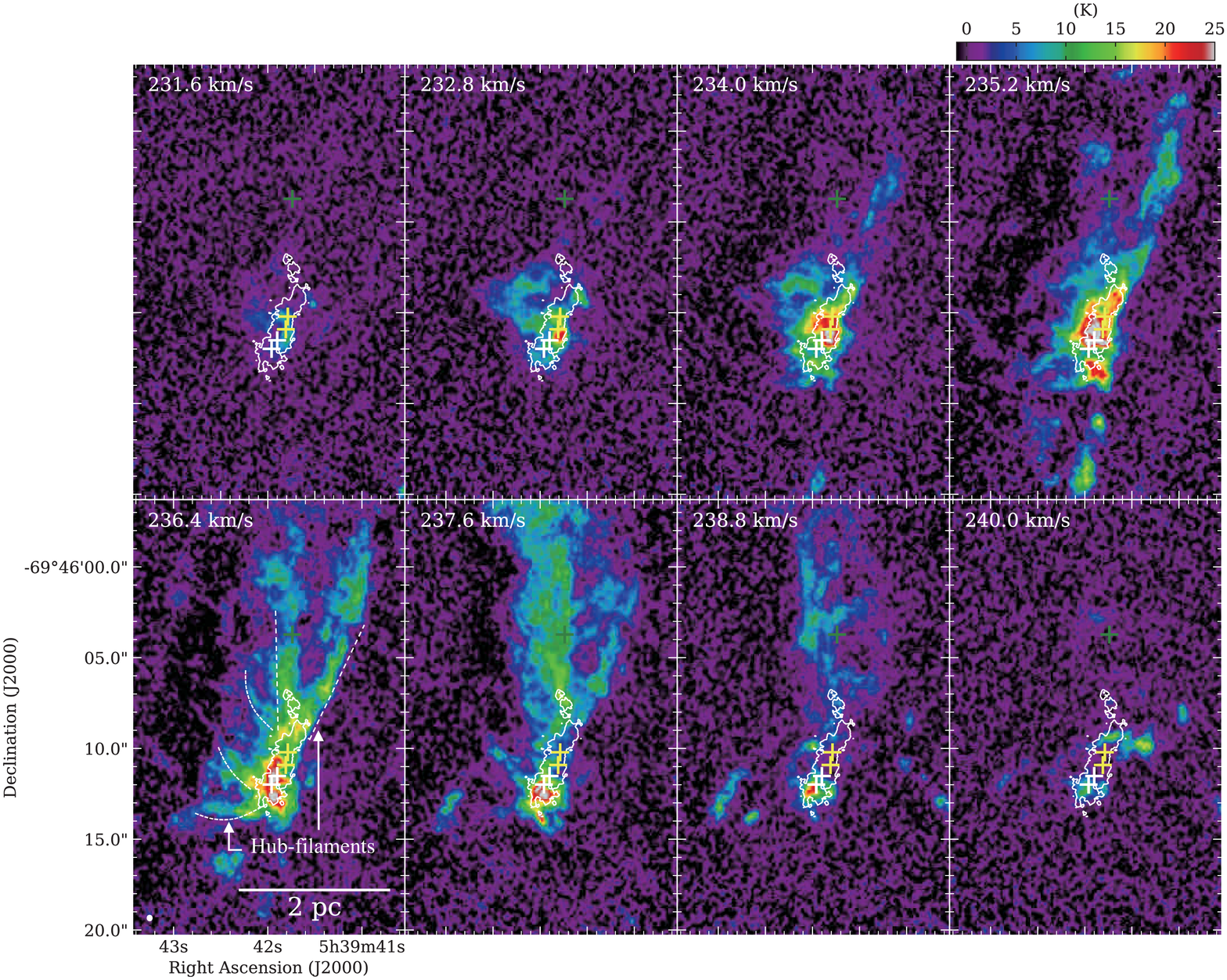}
\caption{
Velocity-channel maps toward the N159W-South clump in $^{13}$CO ($J$ = 2--1). The lowest velocities are given in the upper left corners in each panel. The white contours show the 1.3\,mm continuum emission as shown in Figure \ref{fig:outflow}. The contour level is 0.1\,mJy\,beam$^{-1}$. The angular resolution is given as a white ellipse in the lower left corner of the lower left panel, 0\farcs29 $\times$ 0\farcs25. The white, yellow, and green crosses are the same as those in Figure \ref{fig:outflow}.
\label{fig:chanmap}}
\end{center}
\end{figure}

Figure \ref{fig:13COimage} shows the $^{13}$CO\,($J$ = 2--1) distributions around the N159W-South clump. The overall distributions of both Cycle 1 and 4 data are quite similar, suggesting that the new observations well recover the total flux of the previous one (see Sect. \ref{sec:Obs}). The previous Cycle 1 data represent a relatively simple shape, which is composed of overlapping by the two linear-shaped filaments. In contrast to this, the Cycle 4 data trace further complex substructure of the filaments entangled by the multiple filaments. Looking into the velocity structures as shown in the channel maps of Figure \ref{fig:chanmap}, the complexity becomes more apparent.
For example, one of the most striking features is hub-like filaments entangled toward the dust filament around the velocity channel of $\sim$237\,km\,s$^{-1}$ (dashed lines in the lower left panel in Figure \ref{fig:chanmap}).
Similar morphologies are seen in the Galactic high-mass star-forming clumps \citep[e.g.,][]{Motte07,Peretto13,Williams18}. We calculated the column densities using the $^{13}$CO\,($J$ = 2--1) data assuming the local thermo-dynamical equilibrium. The $``$DisPerSE$"$ \citep{Sousbie11} algorithm was applied to the $^{13}$CO cube data to search the connecting structures (i.e., filaments) in the PPV space (see also the detailed procedure in FTS19). The column density, line mass, and width (FWHM) of the typical filaments are $\sim$10$^{22}$--10$^{23}$\,cm$^{-2}$, a few hundred\,$M_{\odot}$\,pc$^{-1}$, and $\sim$0.1\,pc, respectively, except for the highest column density part with the 1.3\,mm continuum detection. These filaments are significantly more massive than those in the solar neighborhood \citep[see][]{Doris11} and close to those in the high-mass star-forming region (e.g., Vela C, \citealt{Hill12}).

\begin{figure}[htbp]
\begin{center}
\includegraphics[width=190mm]{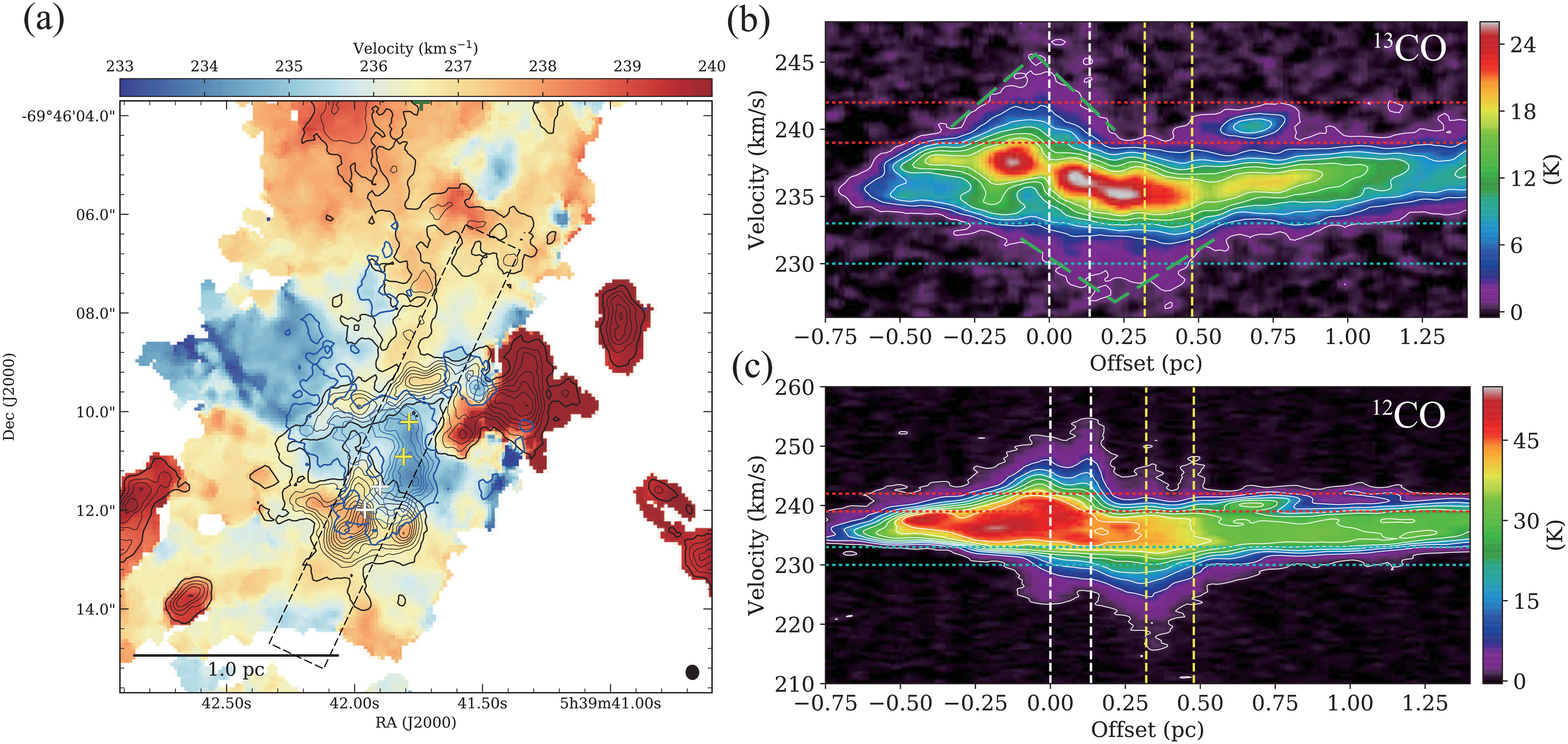}
\end{center}
\caption{
Velocity structures in $^{13}$CO ($J$ = 2--1) toward the N159W-South clump. (a) 
The first-moment intensity-weighted velocity map of $^{13}$CO ($J$ = 2--1) is shown in color scale. 
The blue and black contours show the velocity-integrated intensity of $^{13}$CO ($J$ = 2--1) with a range of 230.0--233.0\,km\,s$^{-1}$ and 239.0--242.0\,km\,s$^{-1}$. The lowest contour level and the subsequent contour step are 3 and 4\,K\,km\,s$^{-1}$, respectively. The white and yellow crosses are the same as those in Figure \ref{fig:outflow}. The angular resolution is given by the ellipse in the lower right corner. (b) A $^{13}$CO ($J$ = 2--1) position-velocity diagram along the regions shown by the dashed rectangle in panel (a). The X-axis represents angular distances in parsecs from the position of MMS-1. The vertical dashed lines in panel (b) represent the position of the millimeter sources with the outflows. The cyan and red horizontal lines show the velocity ranges for the blueshifted and redshifted components, respectively, in panel (a). The green dashed lines stress V-shaped features, possible signs of cloud--cloud collision (see the text). (c) Same as (b) but for $^{12}$CO ($J$ = 2--1).
\label{fig:CCC}}
\end{figure}

The velocity analysis of the $^{13}$CO data found that the blueshifted (230.0--233.2\,km\,s$^{-1}$) and redshifted (239.0--242.0\,km\,s$^{-1}$) components have the complementary distributions (Figure \ref{fig:CCC} (a)). The first-moment intensity-weighted velocity map using the full-velocity range of the $^{13}$CO data also shows a similar trend. Complementary gas distributions are often found in regions with cloud--cloud collision events \citep[e.g,][]{Furukawa09, Matsumoto15, Torii17, Hayashi18, Nishimura18, Sano18, Tokuda18, Fukui17, Fukui18a, Fukui18b}.
\ We made position-velocity (PV) diagrams of $^{13}$CO and $^{12}$CO along the 1.3\,mm continuum filament to investigate the gas kinematics (Figure \ref{fig:CCC}). The $^{13}$CO PV diagram shows two high-velocity components both in the redshifted and blueshifted sides, as indicated by the green dotted lines in Figure \ref{fig:CCC} (b). The positions of the high-velocity (V-shaped) features are close to those of star-forming protostellar sources (MMS-1, 2, 3, and 4). Although the V-shaped structures seem to be related to the protostellar activities, they are not explained by the outflows alone. With respect to the $^{12}$CO PV diagram (Figure \ref{fig:CCC} (c)), the positions of the high-velocity components due to the outflows indeed correspond to those of millimeter sources. On the other hand, there are position discrepancies between the crests of the V-shaped features in $^{13}$CO and those of the outflows in $^{12}$CO. The spatial extent of the V-shaped features is $\sim$0.5\,pc, which is much broader than that of the outflows, $\sim$0.1\,pc. We interpret that the V-shaped features as originating from the cloud--cloud collision as the trigger of the protostar formation in this region (see also discussions in Sect. \ref{d:preCCC}). A similar position discrepancy between a V-shaped feature and an outflow on the PV diagram is also found in the Galactic high-mass star-forming region RCW34 \citep{Hayashi18}. 

\section{Discussions \label{Dis}} 
\subsection{Dynamical state of the massive filament in the N159W-South clump}
In this section, we discuss the formation scenario of the high-mass protostars and the filamentary molecular clouds in the N159W-South region. The line mass of the filaments in this region is as large as $\sim$a few $\times$ 10$^2$--10$^3$\,$M_{\odot}$\,pc$^{-1}$. Such massive filaments should be in a $``$supercritical$"$ state \citep[see][]{Inutsuka97}, and are considered to fragment and radially collapse within the freefall time. We speculate that the massive filaments that we see in the present observations are formed quite recently and may be gravitationally unstable objects (see also discussions in \citealt{Andre16}). The typical velocity dispersion ($\sigma_{\rm v}$) of the N159W-South clump is $\sim$1.7\,km\,s$^{-1}$ in $^{13}$CO, and the resultant virial mass per unit length ($M_{\rm vir, line}$ = 2$\sigma_{\rm v}/G$) is calculated to be 1.3 $\times$ 10$^{3}$\,$M_{\odot}$\,pc$^{-1}$. Note that the magnetic field can be an additional factor to stabilize the filament, depending on the strength as expressed by \cite{Inoue18}. 
In this case, a strong magnetic filed ($\gtrsim$2\,mG) is needed to stabilize the observed massive filament with the line mass of $\sim$2 $\times$ 10$^{3}$\,$M_{\odot}$\,pc$^{-1}$ in the N159W-South clump. This is inconsistent with recent observations toward Galactic massive star-forming regions \cite[e.g.,][]{Pillai15,Pattle18}.
In summary, we suggest that the turbulent and magnetic pressure may not stabilize the filament.

Assuming the cylindrical shape of the filament with a width of $\sim$0.14\,pc (Sect. \ref{R:outflow}), the averaged density is calculated to be $\sim$10$^{6}$\,cm$^{-3}$. The freefall time of the gas density is $\sim$10$^{4}$\,yr, which is consistent with the lifetime of the starless phase of massive dense cores suggested by the observations toward the Galactic massive star-forming regions (\citealt{Motte18} and references therein). The short dynamical time of the outflows ($\sim$10$^{4}$\,yr) indicates the embedded YSOs are in an extremely young phase after their formation. In addition, the presence of an outflow source YSO-SN without the 1.3\,mm continuum detection may imply that the star formation was initiated before the formation of the parental dense material (i.e., dense core), although further careful investigations are needed to obtain a robust conclusion through follow-up observations, as mentioned in the Sect. \ref{R:outflow}. In summary, we suggest that there is a possibility that the filaments and the protostars in the N159W-South region were formed at almost the same time, within an order of $\sim$10$^4$\,yr.
%
%
%
\subsection{Possible formation mechanisms of the massive filament and high-mass protostars \label{d:preCCC}}
\ Recent studies of Galactic star-forming regions suggest that massive filaments (ridges) have hub- or web-like structures, which may be a consequence of the global hierarchical collapse (\citealt{Motte18} and references therein). Such an interpretation is based on the gas kinematics traced by molecular line observations showing a parsec-scale (line-of-sight) velocity gradient along with the filamentary structures \cite[e.g.,][]{Peretto14}. 
In the N159W-South clump, the velocity gradient is small, with a velocity of $\sim$237\,km\,s$^{-1}$ on the top side and a quickly changed velocity of $\sim$235\,km\,s$^{-1}$ close to the positions of the millimeter sources with outflows (yellow crosses in the dotted rectangle of Figure \ref{fig:CCC} (a)).
Although this type of velocity sequence (i.e., positive--negative--positive) can be regarded as gravitational contraction toward the center of the gravitational potential \cite[e.g.,][]{Hacar17} if we directly rely on the 2D velocity map, recent 3D numerical simulation by \cite{Li19} cautioned that such an interpretation based on the line-of-sight velocity map could lead to an incorrect conclusion. They demonstrated that this type of velocity map is reproduced by a collision of the gas flow of two clouds and our velocity map in Figure \ref{fig:CCC} (a) is very similar to the middle panel of Figure 16 in \cite{Li19}. 
With respect to the PV diagram of Figure \ref{fig:CCC} (b), we found two V-shaped features and indicate them with green lines on the blueshifted and redshifted sides around the high-mass star-forming millimeter sources. Numerical simulations \citep{Takahira14} and the synthetic observations \citep{Fukui18b} of colliding clouds also reproduce similar V-shaped structures in the PV diagram. The present gas characteristics are consistent with the previous cases of the cloud--cloud collision as the trigger of massive star formation. Additonally, \cite{Doris18} found a young filament in Taurus with a similar V-shape feature in the PV diagram and they suggested that convergence of flow is a convincing mechanism of the filament formation. In summary, we conclude that the convergent flows or collisions between two clouds are more likely rather than a global hierarchical collapse, to interpret the velocity structure of the $^{13}$CO data in this system.
%
%
%
\subsection{Filament and massive star formation triggered by colliding flows \label{d:CCC}}
\ We previously discussed that a collision between two massive filaments with a line mass of a few 100\,$M_{\odot}$\,pc$^{-1}$ triggered the massive star formation, based on the lower-resolution results (Paper I, see also Figure \ref{fig:13COimage} (b)). However, such a simple model of two colliding filaments may be hard to reconcile with the complex hub filaments revealed by the present observations. The complex filaments themselves are considered to have originated from-much larger scale kinematics, such as colliding flows. Based on the observational study in Taurus, \cite{Tafalla15} proposed a $``${\it fray and fragment}$"$ scenario composed of two steps. The supersonic collision between two flows creates a network of filaments/fibers, and then the filament system fragments into individual cores or groups of cores (chains).  Although the mass of the cores in the N159W-South clump is one or two orders of magnitude higher than that in Galactic low-mass star-forming regions, such as Taurus, there are some common features: the separation of individual cores ($\sim$0.1--0.2\,pc) and its chain-like morphology, as presented by \cite{Tafalla15}.
With respect to the outside of the N159W region, there is another star-forming filament complex in the N159E region located at $\sim$50\,pc away from the N159W-South clump (\citealt{Saigo17}; FTS19). The N159E region has a compact H$\;${\sc ii} region, the Papillon Nebula at the edge/intersection of the hub filament, and thus it is slightly more evolved than the N159W-South clump. Nevertheless, there are little observable differences between the two objects in the filament properties and star-formation activities. The filamentary complexes in both the N159W-South and N159E-Papillon regions have conical shapes with a vertex on the south sides and hub-like features around the high-mass protostars, as well as the massive/dense filaments traced by the 1.3\,mm continuum emission. In the velocity structures of $^{13}$CO (Figures \ref{fig:chanmap}, and \ref{fig:CCC}; see also Figures 2 and 3 in FTS19), there are velocity gradients that are redshifted and blueshifted along the north--south direction, although the degree of the gradient ($\sim$1\,km\,s$^{-1}$\,pc$^{-1}$) in N159W is little smaller than that of the N159E, possibly due to the projection effect. We also found several outflow activities with a dynamical time scale of $\sim$10$^{4}$\,yr along the 1.3\,mm continuum filament in both regions. In summary, the N159W-South clump is supposed to be an almost twin of the N159E-Papillon region.
To explain the synchronized high-mass star and filament formation over a $\sim$50\,pc scale, we consider a large-scale ($>$100\,pc) triggering event rather than the local $\sim$10\,pc motion around the N159W/E regions.
%

A systematic search of large H$\;${\sc i} structures in the LMC \citep{Kim99} found that the N159 region is located along the western edge of a H$\;${\sc i} supergiant shell, SGS19, which has a radius of $\sim$390\,pc and an expansion velocity of $\sim$25\,km\,s$^{-1}$ \citep{Dawson13}. More recently, \cite{Fukui17} pointed out that the molecular ridge, the most massive molecular complex in the LMC, contains the N159 region was formed by large-scale H$\;${\sc i} gas flows, with a relative velocity of $\sim$50\,km\,s$^{-1}$ induced by the last galactic tidal interaction between the LMC and the SMC. The presence of the ongoing large-scale collision is also supported by the distribution of the $A_{\rm v}$ map in this region \citep{Furuta19}. Such a high-velocity H$\;${\sc i} flow could be a promising candidate as the origin of the strong shock compression that produces the massive filaments and massive protostars in the N159W/E regions.

\begin{figure}[htbp]
\plotone{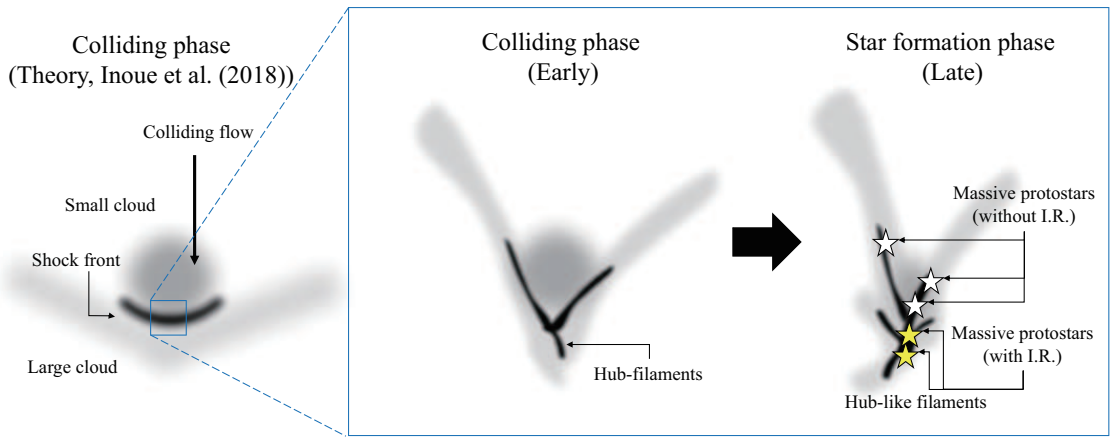}
\caption{
Schematic views of the star formation triggered by a cloud--cloud collision in the N159W-South. A similar figure was presented by FTS19.
\label{fig:CCCponti}}
\end{figure}

\cite{Inoue18} performed numerical simulations of colliding clouds with magnetic field and turbulence and demonstrated that the turbulent inhomogeneous cloud is compressed by the shock wave, and hub filaments are developed within a few 0.1\,Myr after the collision. The first protostar (sink particle) is created in a few $\times$ 10$^4$\,yr after the development of the filaments. A picture of star formation as understood by the theoretical study and the current observations is schematically shown in Figure \ref{fig:CCCponti}. Although there are no large differences in the dynamical time of the outflows toward each protostellar source of more than $\sim$10$^4$\,yrs, the evolutionary stages of the northern three sources (MMS-3, 4, and YSO-SN) are supposed to be younger than the others based on the detection of the infrared emission (Figures \ref{fig:outflow} and \ref{fig:CCCponti}). This distribution is qualitatively explained if the initial small cloud collided with the extended cloud \citep{Inoue18} and the star-formation took place as the propagation of the interaction layer. The evolutionary sequence is also consistent with that in the N159E-Papillon region where the compact H$\;${\sc ii} region is growing at the southern edge of the filamentary clouds (FTS19). 

According to the simulations by \cite{Inoue18}, a large velocity difference between the colliding clouds of the order of 10\,km\,s$^{-1}$ is initially needed to promote high-mass star formation activities (see also \citealt{Inoue13}). However, the density-weighted gas velocity is as small as $\sim$1--2\,km\,s$^{-1}$ with respect to the systemic velocity along the star(sink particle)-forming filament, because the shock speed is decelerated in the dense region when the shock wave induced by the collision hits a dense clump (\citealt{Fukui18a}; see also \citealt{Inoue12,Inoue13}).  
In the N159W-South clump, the mass-weighted velocity (i.e., local peaks on the PV diagram of Figure \ref{fig:CCC} (b)) around the protostellar sources is also not significantly shifted, not more than a few tens of km\,s$^{-1}$, with respect to the systemic velocity, $\sim$236\,km\,s$^{-1}$.
This is consistent with the simulations. We thus suggest that the initial collision velocity might be at least a few $\times$ 10\,km\,s$^{-1}$ based on the H$\;${\sc i} studies as mentioned in this section, and then, the collision shock velocity was considered to be decelerated in the dense region.
In the H$\;${\sc i} colliding flow scenario, we note that there is an issue regarding the formation time-scale of H$_2$ molecules if the large-scale H$\;${\sc i} compression quickly creates the molecular filaments traced by the CO observations. The formation timescale of H$_2$ molecules from H atoms on dust grains is considered to be as long as $\sim$10$^7$ yr, with an atomic gas density of 10$^2$\,cm$^{-2}$ \citep{Hollenbach71,Jura74}, which is much longer than that of the filament formation as discussed above. 
Although it is possible, in principle, to make $\sim$10$^4$\,$M_{\odot}$ molecular filaments after the H$\;${\sc i} gas flow if there is a sufficient mass reservoir around the filamentary cloud \citep{Fukui18a}, further theoretical and observational investigations are needed to clearly understand the relation between the filament formation and high-velocity H$\;${\sc i} flows.

Another interesting feature is that the orientations of the observed outflows are roughly perpendicular to that of the filament (Sect. \ref{R:outflow}). This may indicate that the directions of magnetic fields are also perpendicular to the filament if the outflows were launched along the magnetic field direction. Simulations in \cite{Inoue18} suggest that the magnetic field strength is significantly enhanced at the post-shock layer and massive filaments become perpendicular to the magnetic field. From an observational prespective, for example, \cite{Palmeirim13} revealed that high-density filaments are perpendicular to the magnetic fields, while low-density striations are parallel in low-mass star-forming filaments in Taurus. Recent ALMA observations by \cite{Kong19} toward an infrared dark cloud found that large numbers of CO outflows are preferentially orthogonal to the parental filaments and discussed the presence of strong magnetic field as their origin. Future polarization observations toward the N159W-South clump with ALMA may provide us with further evidence of the cloud--cloud collision as the trigger for massive filament/protostar formation in terms of magnetic field.\\

\section{Summary \label{sec:summary}}
We have carried out ALMA observations with an angular resolution of $\sim$0\farcs25 ($\sim$0.06\,pc) toward the N159W-South region in the LMC. The 1.3\,mm dust continuum traces a clear filamentary feature with a line mass of $\sim$2 $\times$ 10$^3$\,$M_{\odot}$ and it has four local peaks with a strong indication of outflow activities. We have identified a new bipolar outflow source embedded at a $^{13}$CO filament, but it is located at $\sim$2\,pc away from the massive dust filament. We have revealed an early stage of multiple high-mass star formation as a few additional outflow sources along the 1.3\,mm filamentary cloud. The molecular line observations in $^{13}$CO\,($J$ = 2--1) revealed the complex hub structures toward the dust filament, rather than simple linear filaments, as reported in our previous lower-resolution observations. We propose that the massive protostars and filaments are formed from the large-scale flow, which is consistent with the recent theoretical simulations.

\acknowledgments
We thank Doris Arzoumanian and Shu-ichiro Inutsuka for discussions about the filamentary molecular clouds.
This paper makes use of the following ALMA data: ADS/ JAO.ALMA\#2012.1.00554.S and \#2016.1.01173.S. ALMA is a partnership of the ESO, NSF, NINS, NRC, NSC, and ASIAA. The Joint ALMA Observatory is operated by the ESO, AUI/NRAO, and NAOJ. This work was supported by NAOJ ALMA Scientific Research grant No. 2016-03B and JSPS KAKENHI (grant No. 22244014, 23403001, 26247026, 18K13582, 18K13580, and 18H05440). The work of M.S. was supported by NASA under award number 80GSFC17M0002.
\software{CASA (v5.0.0;  \citealt{McMullin07})}



\end{document}